\documentclass[prb,twocolumn,showpacs,preprintnumbers,floatfix]{revtex4-1}

\usepackage{graphicx}
\usepackage{dcolumn}
\usepackage{bm}
\usepackage{latexsym,epsfig}
\usepackage{graphicx}
\usepackage{verbatim}
\usepackage{comment}
\usepackage{amsmath}
\usepackage{amssymb}
\usepackage{stmaryrd}
\usepackage{color}

\newcommand{\beq}{\begin{equation}}
\newcommand{\eeq}{\end{equation}}
\newcommand{\bea}{\begin{eqnarray}}
\newcommand{\eea}{\end{eqnarray}}
\newcommand{\ben}{\begin{eqnarray*}}
\newcommand{\een}{\end{eqnarray*}}
\newcommand{\be}{\begin{enumerate}}
\newcommand{\ee}{\end{enumerate}}
\newcommand{\bfig}{\begin{figure}}
\newcommand{\efig}{\end{figure}}

\begin{document}

\title{Phase diagram of the half-filled one-dimensional $t$-$V$-$V'$ model}

\author{Tapan Mishra, Juan Carrasquilla, and Marcos Rigol} 
\affiliation{Department of Physics, Georgetown University, Washington DC, 20057}

\date{\today}

\begin{abstract}
We study the phase diagram of spinless fermions with nearest- and next-nearest-neighbor 
interactions in one dimension utilizing the (finite-size) density-matrix 
renormalization group method. The competition between nearest- and 
next-nearest-neighbor interactions and nearest-neighbor hopping generates four 
phases in this model: two charge-density-wave insulators, a Luttinger-liquid phase,
and a bond-order phase. We use finite-size scaling of the gap and various structure 
factors to determine the phase diagram.

\end{abstract}


\maketitle

\section{INTRODUCTION}
\label{sect:intro}

The study of quantum phase transitions in strongly interacting systems has become a major 
field of research in condensed matter physics. Quantum fluctuations, rather than thermal 
fluctuations, are responsible for such zero temperature phase transitions.\cite{sachdev_book_01} 
Remarkably, quantum fluctuations play a more dominant role as the dimensionality of the system 
is reduced. In that sense, one and two dimensions are in general an ideal playground where 
the interplay between strong interactions and quantum fluctuations leads 
to the emergence of exotic phases and phenomena. For example, in one dimension, it is known 
that quantum fluctuations melt any order that would otherwise break a continuous 
symmetry.\cite{hohenberg_67,mermin_wagner_66}

In recent years, in addition to traditional condensed matter systems that exhibit reduced dimensionalities, 
ultracold atomic gases in optical lattices have allowed the experimental realization of several strongly 
correlated systems that are of great interest.\cite{bloch_dalibard_review_08} 
The degree of control in those experiments is such that effective one-dimensional regimes are 
now within reach.\cite{cazalilla_citro_11} In addition, the high degree of isolation in optical lattice 
experiments enables the study of nonequilibrium phenomena not accessible in condensed matter 
settings,\cite{kinoshita_wenger_06,trotzky_chen_11} and addressing questions related to the relaxation 
dynamics and thermalization of quantum systems.\cite{rigol_dunjko_08}

One-dimensional systems have been the focus of many recent works on the effects of integrability and 
phase transitions in the dynamics and thermalization of strongly interacting quantum 
systems.\cite{cazalilla_citro_11,polkovnikov10} In particular, one of us (in collaboration with
L. Santos) studied the breakdown of thermalization in the $t$-$V$-$V'$ model, and its relation 
to approaching integrable points and the various insulating phases present in its ground 
state.\cite{rigol_santos_10,santos_rigol_10b} Interestingly, despite the simplicity of the model, 
and the fact that it has been studied in several previous 
works,\cite{emery_noguera_88,hallberg_gagliano_90,nomura_okamoto_94,zhuravlev_katsnelson_97,poilblanc_yunoki_97,schmitteckert_werner_05} 
an accurate phase diagram is not available in the literature.

In this work, we present a detailed study of the phase diagram of the $t$-$V$-$V'$ model. 
For spinless fermions, in its particle-hole symmetric form, the Hamiltonian can be written as 
\begin{eqnarray}
H &=&\sum_{i}\left[-t\,(c_{i}^{\dagger}c_{i+1}^{\phantom \dagger}+\text{H.c.})
+V\left(n_i-\frac{1}{2}\right)\nonumber
\left(n_{i+1}-\frac{1}{2}\right)\right.\nonumber\\
&&\left.+V'\left(n_i-\frac{1}{2}\right)\left(n_{i+2}-\frac{1}{2}\right)\right].
\label{eq:spinless}
\end{eqnarray}
where $c_i^{\dagger}$ and $c_i^{\phantom \dagger}$ are fermionic creation and annihilation operators 
at site $i$, respectively. These operators obey the usual fermionic anticommutation relations 
$\{c_i^{\phantom \dagger},c_j^{\dagger}\}=\delta_{i,j}$. $n_i=c_i^{\dagger}c_i^{\phantom \dagger}$ is the number operator. 
$t$ is the hopping amplitude between neighboring sites, and  $V$ and $V'$ are the 
nearest- and next-nearest-neighbor interactions, respectively. 

We focus our analysis on the half-filled case, i.e., when the number of fermions is equal 
to one-half the number of lattice sites. For $V'=0$, this Hamiltonian can be solved exactly using 
the Bethe ansatz.\cite{cazalilla_citro_11} As $V$ is increased, one finds  a transition from a gapless 
Luttinger liquid (LL) to a gapped charge-density wave  at $V/t=2$. 
For $t=0$, on the other hand, the ground state of model (\ref{eq:spinless}) is very simple 
(a product state) and is determined by the ratio between $V$ and $V'$. For $V'<V/2$ one has a 
$(\ldots~1~0~1~0~1~0~1~0~\ldots)$ charge-density wave (CDW-I) and 
for $V'>V/2$ one has a $(\ldots~1~1~0~0~1~1~0~0~\ldots)$ charge-density wave (CDW-II). Here, 
$1 \, (0)$ denotes the presence (absence) of a fermion at a particular site. However, for finite values of 
$t$, a richer phase diagram emerges.  

Early numerical studies of this model were based on the Lanczos method,\cite{hallberg_gagliano_90} 
for which only small system sizes can be diagonalized. It was found that the competition 
between $t$, $V$, and $V'$ leads to four different phases. Three of those phases have been 
already mentioned, the LL, CDW-I, and CDW-II phases. In addition to those, a bond-order
(BO) phase was also found. This BO phase was somehow missed in some subsequent studies of this 
model.\cite{zhuravlev_katsnelson_97,poilblanc_yunoki_97} More recently, the presence of the 
BO phase was discussed in a density-matrix renormalization group (DMRG) study, but the phase 
diagram was not computed there.\cite{schmitteckert_werner_05} 

It is interesting to note that despite the fact that the physics of the $t$-$V$-$V'$ model is 
well understood,\cite{cazalilla_citro_11} as mentioned before, no accurate phase diagram has been reported 
so far. This is because large finite-size effects make it difficult to determine the precise
boundaries between the phases mentioned above. In this paper, utilizing the DMRG technique, 
we determine the phase diagram of the $t$-$V$-$V'$ model for a broad range of values of $V/t$ and $V'/t$. 
In order to make accurate predictions, we use scaling properties of several quantities, which 
are based on the universality class of the various transitions. Our main result is the phase
diagram reported in Fig.~\ref{fig:phasedia}.

\bfig[!t]
  \centering
  \includegraphics*[width=0.4\textwidth,draft=false]{fig1.eps}
    \caption{(Color online) Phase diagram of the half-filled $t$-$V$-$V'$ model. 
    The phases are, from bottom to top, charge-density wave with a two-site unit cell, 
    Luttinger liquid, bond-order phase, and charge-density wave with a four-site unit 
    cell. The dashed line corresponds to $V'=V/2$, which, in
    the atomic limit, gives the boundary between the CDW-I and CDW-II phases. The hopping amplitude  is set to $t=1$. }
    \label{fig:phasedia}
\efig 

The exposition is organized as follows. In Sec.~\ref{sect:model}, we discuss the model under consideration 
and its formulation in different physical contexts. We also introduce the observables used to describe the 
different ground-state phases. In Sec.~\ref{sect:gaplesstogapped}, we provide details on the different 
approaches used to detect the transitions from gapless to gapped phases. In particular, we determine the 
transition line from the LL to the CDW-I as well as the transition from the LL to the BO phase. 
In Sec.~\ref{sect:gappedtogapped}, the transition from the BO phase to the CDW-II is studied by performing 
finite-size scaling analysis of the structure factor. We also compare the obtained phase diagram with 
the one found in previous studies. Finally, in Sec.~\ref{sect:conc}, we briefly summarize our main conclusions.

\section{Model, observables, and approach}
\label{sect:model}

The $t$-$V$-$V'$ model for spinless fermions [Eq.~\eqref{eq:spinless}] can be mapped onto 
two other models of much interest in condensed matter and cold gases. 

The first of those two models is a well-known spin chain. Using the Jordan-Wigner 
transformation,\cite{jordan_wigner_28}
\begin{eqnarray}
&&c^{\dag}_i=S^{+}_i \prod^{i-1}_{\beta=1}e^{i\pi (S^z_{\beta}+\frac{1}{2})}, \quad
c_i^{\phantom \dagger}=\prod^{i-1}_{\beta=1}e^{-i\pi (S^z_{\beta}+\frac{1}{2})} S^-_i\nonumber\\
&&c_i^{\dagger}c_i^{\phantom \dagger} =S^z_i+\frac{1}{2},
\end{eqnarray} 
where $S_i^x$,$S_i^y$,$S_i^z$ are spin-$\frac{1}{2}$ operators at site $i$,
the Hamiltonian \eqref{eq:spinless} takes the form
\begin{equation}
H=\sum_i\left[-2t\,(S_i^xS_{i+1}^x+S_i^yS_{i+1}^y)+VS_i^zS_{i+1}^z+V'S_i^zS_{i+2}^z\right].
\label{eq:hei}
\end{equation}
This is the spin-1/2 $XXZ$ chain with an additional next-nearest-neighbor 
$S^zS^z$ interaction term. For the mapping, we have assumed 
that the system has open boundary conditions, as will be the case throughout this work.
For periodic boundary conditions, a boundary term appears which depends on the total number of
fermions (total $S^z$) in the chain.\cite{lieb_shultz_61}

Furthermore, one can map the spin model above onto a model of impenetrable (hard-core) bosons,
known as the lattice Tonks-Girardeau gas within the cold-gases community.\cite{cazalilla_citro_11}
For this, one can use the Holstein-Primakoff transformation for spin-1/2 
particles.\cite{holstein_primakoff_40}
\begin{eqnarray}
&&S_i^+=a_i^{\dagger}\sqrt{1-a_i^{\dagger}a_i^{\phantom \dagger}},\quad  
S_i^-=\sqrt{1-a_i^{\dagger}a_i^{\phantom \dagger}}~a_i^{\phantom \dagger}\\
&&S_i^z=a_i^{\dagger}a_i^{\phantom \dagger}-\frac{1}{2}.\nonumber
\end{eqnarray}
Under this transformation, Eq.~(\ref{eq:hei}) can be written as
\begin{eqnarray}
H &=&\sum_{i}\left[-t\,(a_{i}^{\dagger}a_{i+1}^{\phantom \dagger}+\text{H.c.})
+V\left(n^b_i-\frac{1}{2}\right)\nonumber
\left(n^b_{i+1}-\frac{1}{2}\right)\right.\nonumber\\
&&\left.+V'\left(n^b_i-\frac{1}{2}\right)\left(n^b_{i+2}-\frac{1}{2}\right)\right].
\label{eq:ham}
\end{eqnarray}
where $a_i^{\dagger}\,(a_i^{\phantom \dagger})$ is the bosonic creation (annihilation) operator obeying the bosonic 
commutation relations $[a_i^{\phantom \dagger},a_j^\dagger] = \delta_{i,j}$ and $n^b_i = a_i^{\dagger}a_i^{\phantom \dagger}$ is the bosonic 
number operator. One also has an additional constraint $a_i^{{\dagger}2}=a_i^{2 \phantom \dagger}=0$ that precludes multiple
occupancies of the lattice sites. 

From the derivations above, one can see that spinless fermions, spins, and hard-core bosons share 
the same spectrum, and diagonal (density and $S^z$) correlations. Hence, the phase diagram obtained 
in this study for spinless fermions is also relevant to the corresponding spin and bosonic models. 
Experimentally, hard-core bosons have been realized in one-dimensional geometries in the 
presence\cite{paredes_widera_04} and absence\cite{kinoshita_wenger_04} of a lattice.

In order to compute the ground-state properties of the $t$-$V$-$V'$ model, we use the finite-size 
DMRG algorithm with open boundary conditions.\cite{white_92,schollwock_review_05} This method is best 
suited for (quasi-)one-dimensional problems and has been extensively used to study
quantum spin chains.\cite{schollwock_review_05} To minimize finite-size effects and 
obtain accurate extrapolations, we study systems with up to 700 sites (1000 in fewer cases) 
retaining $128$ density-matrix eigenstates. The weight of the states discarded in the density 
matrix is less than $10^{-6}$ in all cases. To improve the convergence, at the end of each
DMRG step, we use a finite-size sweeping procedure.\cite{schollwock_review_05} 

To characterize the various phases of this model, we have studied several quantities. 
Here, we report results for the single-particle excitation gap,
\beq
G_L=E(L,N+1)+E(L,N-1)-2E(L,N).
\label{eq:gap}
\eeq
It allows us to distinguish gapped and gapless phases. In Eq.~\eqref{eq:gap}, 
$E(L,N)$ is the ground-state energy of a system with $L$ sites and $N$ fermions.

To understand the transition to the CDW phases, we calculate the structure factor,
which is the Fourier transform of the density-density correlation function
\begin{equation}
 S(k)=\frac{1}{L^2}\sum_{i,j}{e^{ik(i-j)}(\langle{n_{i}n_{j}}\rangle-\langle n_i\rangle\langle n_j
\rangle)}.
\label{eq:str}
\end{equation}

Finally, the transition to the BO phase can be identified by calculating the bond-order 
parameter
\beq
O_{BO}=\frac{1}{L}\sum_i(-1)^i B_i,
\label{eq:obow}
\eeq
where
\beq
B_i=\langle c_i^\dagger c_{i+1}^{\phantom \dagger}+c_{i+1}^\dagger c_i^{\phantom \dagger}\rangle.
\eeq

In the remainder of the paper, we set $t=1$.

\section{Gapless to gapped phase transitions}
\label{sect:gaplesstogapped}

We first focus on the transition between the gapless LL phase and the 
two gapped phases that surround it, which are the CDW-I and the BO phases. These two transitions 
can be understood using LL theory, from which one obtains that the Luttinger 
parameter is $K=1/2$ at the transition points.\cite{cazalilla_citro_11} One can then use this 
knowledge to determine the boundaries between the LL and CDW-I/BO phases. The idea would be to 
compute $K$ using the decay of correlations for finite systems, make an extrapolation to the 
thermodynamic limit, and then find the values of $V'$ that for a given value of $V$ 
result in $K=1/2$. This approach was used, for example, to calculate the phase diagram of 
the extended Hubbard model.\cite{kuhner_white_00} We find that such a procedure  
leads to inconclusive results for the $t$-$V$-$V'$ model. The selection of the range of 
distances used to fit $K$ from correlation functions, such as the one-particle density matrix 
$\rho_{ij}=\langle c^{\dagger}_ic_j^{\phantom \dagger}\rangle$, leads to a wide range of values of 
$K$ for any given finite system. After extrapolation to the thermodynamic limit, the errors in 
the determination of the critical $V'$, for each value of $V$, were found to be
large. One could also try to use the appropriate functional form of the relevant 
correlation functions as obtained in low-energy effective theories, like it has been done 
in the context of spin chains\cite{vekua_honecker_07,hikihara_furusaki_04} and
zigzag ladders,\cite{hikihara_momoi_10} but we have followed a different approach.

Our approach is based on the study of the closing of the single-particle excitation gap 
$G_L$ [see Eq.~\eqref{eq:gap}] when entering the LL phase. $G_L$ can be accurately determined 
within DMRG at a relatively low computational cost 
(in particular, if one compares it with the cost of computing 
correlation functions). Of course, $G_L$ is finite for any finite system even in phases that are 
gapless in the thermodynamic limit. As an example, in Fig.~\ref{fig:LGV} we show $LG_L$ as a function of
$V'$ and fixed $V=4$, for two different intervals of $V'$, and various system sizes. One can see 
there that $LG_L$ is finite for all values of $V'$ and all system sizes. However, in some parameter 
regimes $LG_L$ does not change with increasing system size, i.e., $G_L\sim 1/L$ 
(vanishes in the thermodynamic limit), while in other parameter regimes, $LG_L$ increases 
with increasing system size and so $G_L$ is finite in the thermodynamic limit.

\bfig[!t]
  \centering
  \includegraphics*[width=0.45\textwidth,draft=false]{fig2.eps}
    \caption{(Color online) Scaling of the gap $LG_L$ plotted as a function of $V'$ for $V=4.0$. 
The coalescence of different curves for $V' \gtrsim 1.0$ in (a) and $V'\lesssim 2.9$ (b) indicates the 
transition from gapped to gapless phase.}
    \label{fig:LGV}
\efig

One can then extrapolate $G_L$ to the thermodynamic limit by fitting it to a polynomial in terms of $1/L$ 
and obtain the value of $G_{L\rightarrow \infty}$ by varying $V'$ for all the values of $V$. 
Results for the extrapolated gap $G_{L\rightarrow \infty}$ as a function of $V'$ for $V=4$ are shown 
in Fig.~\ref{fig:gapther}. This plot shows a clear transition from a gapped to a gapless phase and 
then to a gapped phase. The system becomes gapless at $V'\sim 1.0$ and the gap reopens at $V'\sim 2.9$. 
Still, within this approach, one encounters the difficulty of pinpointing the exact values of $V'$ for 
the phase transition.

\bfig[!t]
  \centering
  \includegraphics*[width=0.4\textwidth,draft=false]{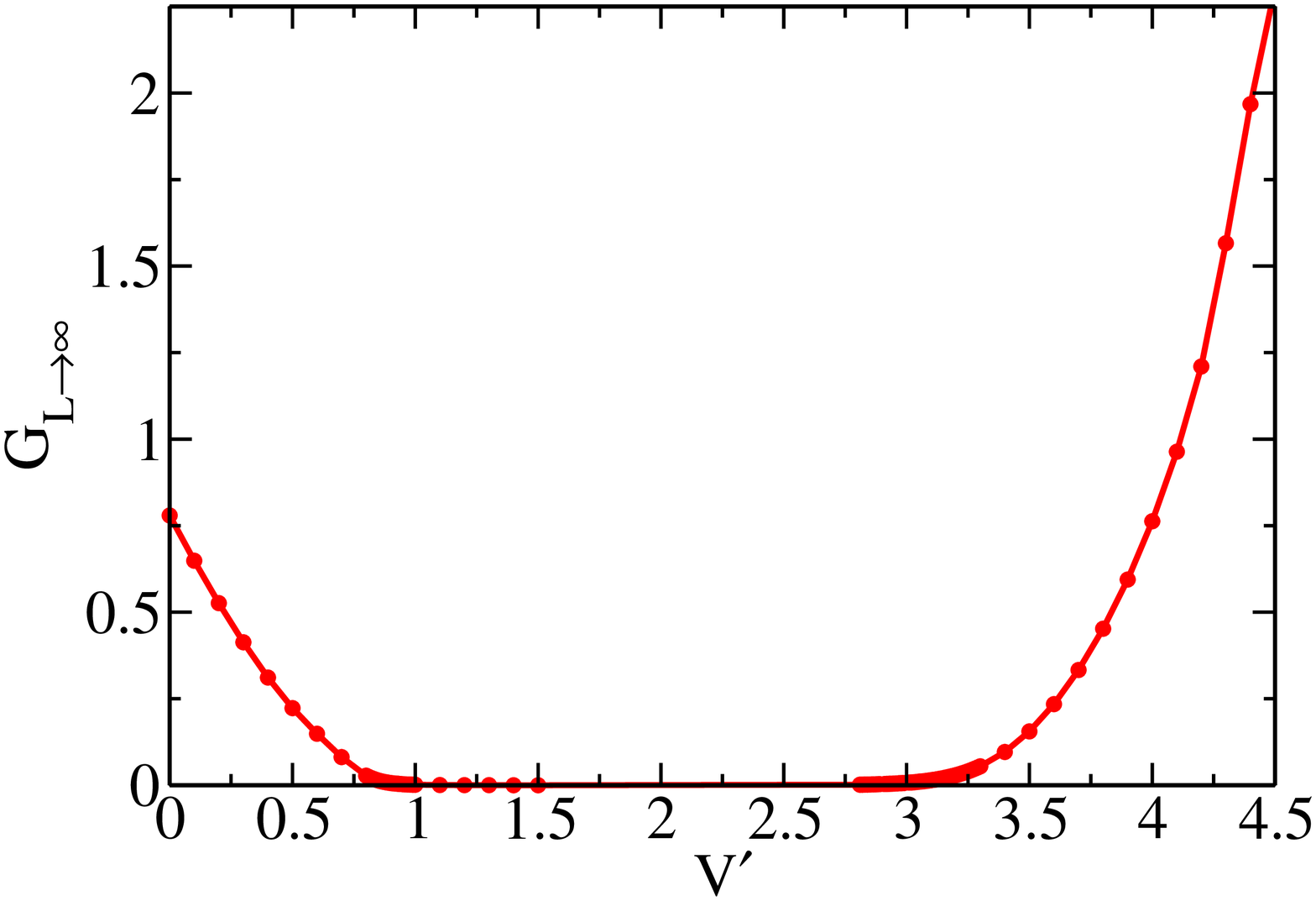}
    \caption{The extrapolated gap in the thermodynamic limit is plotted as a function of $V'$ for $V=4$. 
    There is a transition from a gapped to a gapless phase at $V' \sim 1.0$ and then to a 
    gapped phase again at $V' \sim 2.9$. }
    \label{fig:gapther}
\efig

In order to obtain accurate values of $V'$ for the boundaries between the gapped and gapless phases,
we make use of the knowledge that the transition between the gapless and gapped phases belongs to the
Berezinskii-Kosterlitz-Thouless (BKT) type.\cite{cazalilla_citro_11} At the BKT transition the 
gap closes as 
\begin{eqnarray}
G\sim \exp\left[-\frac{a}{\sqrt{|V'-V'_c|}}\right],
\label{eq:bkt}
\end{eqnarray}
where $a$ is a constant. 


The correlation length $\xi$, which is closely related to the gap, is finite in the gapped phase, diverges 
at the critical point as $\xi\sim G^{-1}$, and remains infinite in the LL phase. We utilize the following 
finite-size-scaling relation for the gap in the vicinity of the phase transition,
\begin{equation}
L G_L \times \left(1+\frac{1}{2\ln{L}+C} \right)= F\left( \frac{\xi}{L} \right),
\label{eq:scaling}
\end{equation}
where $F$ is a scaling function and $C$ is an unknown constant to be determined. This scaling ansatz 
resembles the analogous scaling relation for the resistance in the charge-unbinding transition 
of the two-dimensional Coulomb gas.\cite{wallin_weber_95} In such a transition, which is also of the BKT type,
the resistance near the critical point behaves as the gap does in the gapped to LL transition; namely, 
it vanishes exponentially as in Eq.~\eqref{eq:bkt}. 

\bfig[!b]
  \centering
  \includegraphics*[width=0.45\textwidth,draft=false]{fig4.eps}
    \caption{(Color online) The scaled gap $L G_L'$ as function of the $x_L=\ln L -a/\sqrt{|V'-V'_c|}$.
     The insets show the scaled gap as function of $V'$. (a) Transition from the CDW-I to the LL phase.
      (b) Transition from the LL to the BO phase.  Both transitions are presented
           for a fixed value of $V=4$.
     } 
    \label{fig:LGVscale}
\efig

At the critical point and close to it within the LL phase, one expects the values of $F(\xi/L)$ to be 
system-size independent because of the divergence of the correlation length; i.e., 
plots of $LG_L'=LG_L \left[1+1/\left(2\ln{L}+C\right) \right]$ as function 
of $V'$ for different system sizes should merge in that region. Furthermore, the resulting curves of 
plotting  $L G_L'$ as a function of $\xi/L$ for several values of $L$ should also be system-size 
independent. Equivalently, one can plot $L G_L'$ as a function of $x_L=\ln L -\ln \xi$ to get the collapse 
of the curves. In our calculations, the values of $a$, $C$, and $V'_c$ are fitted to produce the best 
possible collapse of the data within the gapped side of the phase transition. On that side, we take into 
account that the correlation length diverges as $\xi\sim \exp \left[ a/\sqrt{|V'-V'_c|} \right]$. We 
have tested the accuracy of this procedure by determining the critical value of $V$ for the transition 
between the LL and CDW-I phases when $V'=0$. As mentioned in the introduction, the critical value of $V$ 
in this case is known analytically (the Heisenberg point). We considered systems with up to 700 sites and 
obtained a critical value $V_c=2.02 \pm 0.01$, which is consistent with the analytical result. 

When $V'\neq 0$, this model is not exactly solvable. In Fig.~\ref{fig:LGVscale}, we show results for 
$L G_L'$ vs~$x_L$ (main panels), as well as $L G_L'$ vs~$V'$ (insets), for $V=4$ and several values of 
$L$. The main panels make evident the collapse of all the data for the rescaled gap as a function of 
$x_L$, shown here only within the gapped phases. Two important limits can be understood by analyzing the 
collapse curves. As the critical point is approached ($x_L\to -\infty$), the scaling function approaches 
a constant value, which in turn, implies the vanishing of the gap. In the limit of large $x_L$, one can 
see that the scaling function increases rapidly, which is necessary for the gap to be finite in the gapped 
phases. We then confirm two aspects of the BKT transition, the exponentially divergent correlation length 
and the logarithmic corrections to the gap. The insets in Fig.~\ref{fig:LGVscale} clearly show that, when 
coming from the gapped phases, the $L G_L'$ curves merge at a point [$V'_c=1.16$ for the CDW-I to LL 
phase, Fig.~\ref{fig:LGVscale}(a), and $V'_c=2.55$ for the BO to LL phase, Fig.~\ref{fig:LGVscale}(b)] 
and remain close to each other for some finite region within the gapless phase. Moreover, as the size of 
the system increases, the interval over which the curves are seen to collapse within the LL phase 
increases. As said before, because of the divergence of the correlation length in the LL phase, the 
curves should be system-size independent in the vicinity of the transition point, where the scaling 
relation \eqref{eq:scaling} holds. 

\subsection{CDW-I to LL transition}\label{cdwi}

As mentioned in the introduction, only two phases are present in the system for $V'=0$. Those are a LL phase
for $0 < V < 2$ and a CDW-I phase for $V > 2$. By adding a small $V'$, the transition between LL to the CDW-I
is shifted to larger values of $V$, thus enhancing the stability of the LL phase. This is because the interplay 
between CDW configurations of the type $(\ldots~1~0~1~0~1~0~\ldots)$, favored by increasing values of $V$, competes 
with those of the type $(\ldots~1~1~0~0~1~1~0~0~\ldots)$, favored by increasing values of $V'$, thus making the system 
gain kinetic energy because of the net reduced charge ordering effects. Nevertheless, for large enough values of $V$, 
the CDW-I phase is prevalent. In order to see that this is the case, we have calculated the 
density-density structure factor $S(k)$ as defined by Eq.~\eqref{eq:str}. Because of the charge order, the 
CDW-I phase is characterized by a finite value of $S(\pi)$ in the thermodynamic limit. In 
Fig.~\ref{fig:linvstr1}, we show $S(\pi)$ as function of $1/L$ for different values of $V'$, and for $V=4$. 
It is apparent in that figure that, for $V'\lesssim 1.0$, $S(\pi)$ is finite in the thermodynamic 
limit. For $V'\gtrsim 1.2$, on the other hand, $S(\pi)$ becomes very small as the system size increases, 
which suggests that it will be zero in the thermodynamic limit. This is consistent with the results from 
the scaling of the gap that, for $V=4$, predicted the critical value 
of $V'$ for the transition between the LL and the CDW-I phase to be $V'=1.16$. Despite the fact that with
the finite-size scaling of the structure factor $S(\pi)$ it is difficult to locate the exact transition point, 
this calculation provides evidence of the nature of the CDW-I phase and the way the charge order develops as 
one crosses the critical region. 

\bfig[!t]
  \centering
  \includegraphics*[width=0.4\textwidth,draft=false]{fig5.eps}
    \caption{(Color online) Density-density structure factor $S(\pi)$ as a function of $1/L$ for different values 
of $V'$, for $V=4$. $S(\pi)$ is predicted to be finite in the thermodynamic limit for 
$V'\lesssim 1.2$, making evident the existence of a CDW-I phase in that region of parameter space.}
    \label{fig:linvstr1}
\efig

\subsection{LL to BO transition}\label{bo}

At the other boundary of the LL phase, when $V'>V/2$, there is an instability toward the formation 
of a gapped BO phase.\cite{emery_noguera_88,hallberg_gagliano_90} In the limit of vanishing $V$, the bond-order 
phase arises because of the competition between the kinetic energy term and the next-nearest-neighbor interaction 
that induces a CDW-II order. A finite value of $V$ competes with the bond ordering induced by $V'$, thus enhancing the 
stability of the LL phase and increasing the critical value of $V'$. As mentioned previously, the BO phase is 
characterized by a finite value of $O_{BO}$ [see Eq.~\eqref{eq:obow}] in the thermodynamic 
limit. Although this phase exhibits bond oscillations as unveiled by the $O_{BO}$ parameter, the system exhibits no 
charge order. Therefore, two quantities can be used to characterize the BO phase, the gap and $O_{BO}$, 
which are both finite in that phase. In a finite-size system, however, BO oscillations are present in the LL phase, 
but vanish as the system size in increased.

\bfig[!t]
  \centering
  \includegraphics*[width=0.42\textwidth,draft=false]{fig6.eps}
    \caption{(Color online) Finite-size scaling of $O_{BO}$ for different values of $V'$, for $V=4$, across the LL to BO phase.}
    \label{fig:obow}
\efig
 
Following the arguments in Ref.~\onlinecite{ejima_nishimoto_07} and~\onlinecite{white_affleck_scalapino_02}, the strength 
of such oscillations is assumed to decay as $L^{-K}$. This scaling relation holds inside the LL region and in particular
at the transition point between the LL and the BO phase, where the Luttinger-liquid parameter takes the value $K=1/2$.
Because of that, we extrapolate the BO parameter using $L^{-1/2}$ rather than $L^{-1}$. In Fig.~\ref{fig:obow}, we 
plot $O_{BO}$ as a function of $L^{-1/2}$ for different values of $V'$ at $V=4$. One can see there that $O_{BO}$ 
extrapolates to a nonzero value in the thermodynamic limit for $V'\gtrsim 2.6$. Despite the large size effects 
that are present at the BKT transition, for values of $V'\lesssim 2.5$, $O_{BO}$ extrapolates to 
very small values. These results are compatible with the critical value $V'_c=2.55$ obtained from the finite-size-scaling 
analysis of the opening of the charge gap. Note that in Fig.~\ref{fig:obow}, as the values of $V'$  decrease and approach 
the critical value $V_c$, the curves become closer to straight lines. This supports our assumption that, at the transition 
point, the BO oscillations decay as $L^{-1/2}$.

\section{Gapped to gapped phase transition}
\label{sect:gappedtogapped}

\subsection{BO to CDW-II transition}

Ultimately, when $V'$ is very large, the system always forms a CDW-II insulator as the energy 
of the configuration (\ldots~1~1~0~0~1~1~\ldots) becomes energetically more favorable. Hence, 
within the gapped region with $V'>V/2$, there is an additional phase transition between the BO 
and CDW-II phases. The latter phase is characterized by a finite value of $S(\pi/2)$ in
the thermodynamic limit.

Taking $S(\pi/2)$ as the order parameter for the  CDW-II phase, we obtain the critical point 
accurately by means of scaling theory. We start with the ansatz
\beq
S(\pi/2)L^{2\beta/\nu}=F\left( \left(V'-V'_c \right)L^{1/\nu} \right)
\eeq
where $\beta$ is the critical exponent corresponding to the universality class of the phase 
transition and $\nu$ is the correlation exponent. $F$ is a scaling function and $V'_c$ is now the critical 
point for transition between the BO to CDW-II phases. 

\bfig[!t]
  \centering
  \includegraphics*[width=0.4\textwidth,draft=false]{fig7.eps}
    \caption{(Color online) Scaled $S(\pi/2)$ vs the scaled control parameter  for $V=4$. The collapse of all 
    the data points onto a single curve confirms the critical point at $V'_c=4.293$. Inset: Crossing of 
the scaled $S(\pi/2)$ for $V=4$ and different system sizes. All the curves intersect at $V'_c=4.293$ 
indicating the transition point.}
    \label{fig:figcdw2b}
\efig

Comparing results for different system sizes as $V'$ is changed, we found that the curves for 
different system sizes intersect at a unique critical point if $\beta=0.125$ and $\nu=1$ (these 
are the exponents of the 2D Ising universality class). Results for a value of $V=4$ 
are shown in Fig.~\ref{fig:figcdw2b}, for which $V'_c\simeq 4.293$. The scaled $S(\pi/2)$ 
as a function of scaled control parameter $V'$ is plotted in Fig.~\ref{fig:figcdw2b}. 
The collapse of the data confirms the accuracy of $V'_c$ as determined from the crossing in 
the inset of Fig.~\ref{fig:figcdw2b}. Notice that, as $V'$ is increased, the transition between 
the BO to CDW-II occurs while the charge gap continues to grow monotonically, i.e., it does 
not vanish at the transition point. See the behavior of the charge gap in Fig.~\ref{fig:gapther} 
for values around the critical $V'_c\simeq 4.293$.
 
\bfig[!h]
  \centering
  \includegraphics*[width=0.4\textwidth,draft=false]{fig8.eps}
    \caption{(Color online) Scaling of $S(\pi/2)$ vs $1/L$ for $V=4$ and different values of $V'$ between $V'=4.25$ 
and $V'=4.32$ in steps of $0.01$. One can see that $S(\pi/2)$ extrapolates to zero below the critical
value $V'_c=4.293$.}
    \label{fig:figspi2}
\efig

In order to have an independent check that the previous approach leads to the correct results, 
we have also extrapolated the value of $S(\pi/2)$ to thermodynamic limit. The results from this 
procedure, once again for $V=4$, are presented in Fig.~\ref{fig:figspi2}. There, we plot $S(\pi/2)$ 
vs $1/L$ for different values of $V'$. The values of the extrapolated $S(\pi/2)$ clearly converge 
to zero for $V<V'_c$. 

A complete analysis like the one presented so far, but for different values of 
$V$, allowed us to determine the phase diagram presented in Fig.~\ref{fig:phasedia}. 

\bfig[!h]
  \centering
  \includegraphics*[width=0.4\textwidth,draft=false]{fig9.eps}
    \caption{(Color online) The same phase diagram as in Fig.~\ref{fig:phasedia} but now in the $t/V$ vs $V'/V$ plane.}
    \label{fig:phasedia2}
\efig

Finally, to allow for a direct comparison with early calculations for this model, reported in 
Refs.~\onlinecite{emery_noguera_88,hallberg_gagliano_90}, we present in Fig.~\ref{fig:phasedia2} 
the phase diagram in the plane $t/V-V'/V$. This phase diagram is in qualitative
agreement with the one reported in Refs.~\onlinecite{emery_noguera_88,hallberg_gagliano_90}. 
However, quantitative differences on the precise location of the phase boundaries are apparent. 
Although we did not study the region of the phase diagram where $t/V\to 0$, the trend of our 
transition lines suggests that they will cross for small but finite values of $t/V$, implying a 
tricritical point.\cite{emery_noguera_88,hallberg_gagliano_90}

\section{Conclusion}
\label{sect:conc}
We have presented a comprehensive study of the phase diagram of the $t$-$V$-$V'$ model in one dimension.
Using the scaling of the gap in the BKT transition between the LL phase and the CDW-I/BO phases, 
we have obtained accurate results for the boundaries between the gapless and gapped phases of 
this model, which are confirmed by the extrapolation of various order parameters. The phase transition 
between the BO phase and the CDW-II phase (both of which are gapped) was also studied using scaling theory
for the density-density structure factor. This latter phase transition was found to be much less sensitive 
to finite-size effects and the exponents computed were found to be consistent with those of the two-dimensional 
Ising universality class. 

\section{Acknowledgments}
This work was supported by the Office of Naval Research. We thank M. A. Cazalilla, A. Muramatsu, 
E. Orignac and R. V. Pai for useful discussions. We also thank C. Varney, E. Khatami, and E. Malatsetxebarria 
for useful comments on the manuscript.

\end{document}